\documentclass[12pt,a4paper,final]{iopart}

\usepackage{iopams}  
\usepackage{graphicx}
\usepackage{color}
\usepackage[breaklinks=true,colorlinks=true,linkcolor=blue,urlcolor=blue,citecolor=blue]{hyperref}

\begin{document}

\title[T. Kr\"ahenmann: Fermi edge singularities]{Fermi edge singularities in transport through lateral GaAs quantum dots}

\author{Tobias Kr\"ahenmann$^{1}$, Livio Ciorciaro$^{1}$, Christian Reichl$^{1}$, Werner Wegscheider$^{1}$, Leonid Glazman$^{2}$, Thomas Ihn$^{1}$, Klaus Ensslin$^{1}$}
\address{$^1$Solid State Physics Laboratory, ETH Z\"urich, CH-8093 Z\"urich, Switzerland}
\address{$^2$ Department of Physics, Yale University, New Haven, Connecticut 06520, USA}
\ead{tobiaskr@phys.ethz.ch}

\begin{abstract}
We measure tunnelling currents through electrostatically defined quantum dots in a GaAs/AlGaAs heterostructure connected to two leads. For certain tunnelling barrier configurations and high sample bias we find a pronounced resonance associated with a Fermi edge singularity. This many-body scattering effect appears when the electrochemical potential of the quantum dot is aligned with the Fermi level of the lead less coupled to the dot. By changing the relative tunnelling barrier strength we are able to tune the interaction of the localised electron with the Fermi sea.
\end{abstract}

\submitto{\NJP}
% \maketitle

\section{Introduction}

Coulomb interaction of conduction electrons in semiconductor heterostructures leads to a variety of many-body phenomena, such as fractional quantum Hall ground states \cite{tsui_two-dimensional_1982,heinonen_composite_1998}, Kondo correlations \cite{ng_-site_1988,glazman_resonant_1988,jones_low-temperature_1988,goldhaber-gordon_kondo_1998} and Fermi edge singularities \cite{matveev_interaction-induced_1992,geim_fermi-edge_1994,benedict_fermi_1998,thornton_many-body_1998,hapke-wurst_magnetic-field-induced_2000,frahm_fermi-edge_2006,skolnick_observation_1987,cobden_finite-temperature_1995,ruth_fermi_2008}. Fermi edge singularities have first been theoretically predicted for X-ray absorption in metals \cite{nozieres_singularities_1969,mahan_excitons_1967} and have been adapted to the case of electron tunnelling through an impurity (quantum dot) \cite{matveev_interaction-induced_1992}. The first experimental observation of the Fermi edge singularity in electron tunnelling \cite{geim_fermi-edge_1994} was followed by intensive studies in a magnetic field \cite{benedict_fermi_1998,thornton_many-body_1998,hapke-wurst_magnetic-field-induced_2000,frahm_fermi-edge_2006}. Common to all these experiments is the vertical alignment of the tunnelling contacts in MBE-grown barrier structures, i.e. the tunnelling current was perpendicular to the heterostructure layers. On the one hand this brings the localised state spatially close to the Fermi sea, thus increasing the interaction strength, but on the other hand this prevents tuning tunnelling barrier strengths. Given this wealth of experiments on Fermi Edge singularities and the extensive research on laterally defined quantum dots in GaAs based heterostructure it is surprising that Fermi edge singularities have not been consistently reported and investigated in these structures. We are aware of one unpublished result \cite{zumbuhl__????}.\\
The Fermi edge singularity is due to the Coulomb interaction of a localised electron with the continuum of a Fermi sea. A polaron-like virtual state created in the course of tunnelling \cite{mahan_many-particle_2000} enhances the tunnelling amplitude for electrons close to the Fermi level and results in a singular behaviour of the tunnelling current \cite{matveev_interaction-induced_1992}. This singularity is cut off by the finite lifetime of the occupied resonant state. In the zero-temperature limit one finds \cite{matveev_interaction-induced_1992}
\begin{equation}\label{formula_1}
I\propto \sqrt{\left[(\varepsilon_\mathrm{F}-\varepsilon)^2+\Gamma^2\right]}^{-\beta}\times\left(\frac{\pi}{2}+\arctan \left(\frac{\varepsilon_\mathrm{F}-\varepsilon}{\Gamma}\right)\right).
\end{equation}
Here, $\varepsilon_\mathrm{F}$ is the Fermi energy, i.e. the position of the resonance, $\Gamma$ denotes the width of the resonance due to the finite lifetime of the electron on the localised state. The exponent $\beta$ is related to the scattering phase shift $\delta$ of the tunnelling electron and thus the interaction strength of the localised state with the screening Fermi sea.\\
In the case of a single scattering channel (e.g. one spin-polarized edge channel) and small $\beta$ we get $\beta\approx2\delta/\pi$. Using Friedel's sum rule we find $Q/e=\delta/\pi=\beta/2$, where $Q/e$ is the fraction of charge screened by the lead, which by definition is the leverarm $\alpha_\mathrm{S}$ of the source lead (i.e. the Fermi reservoir which exhibits the singularity) on the quantum dot. Thus we expect values of $\beta = 2\cdot\alpha_\mathrm{S}$ of a few ten percent.\\
Matveev and Larkin \cite{matveev_interaction-induced_1992} treat the case of a very asymmetrically coupled quantum dot and predict the singularity to appear due to the interaction with the lead which is less coupled. The singularity will be smeared out due to the finite lifetime of the state, which is dominated by the lead which is more strongly coupled.

\section{Experiment}

\begin{figure}[t!hb!]
 \centering
 \includegraphics[width=\textwidth]{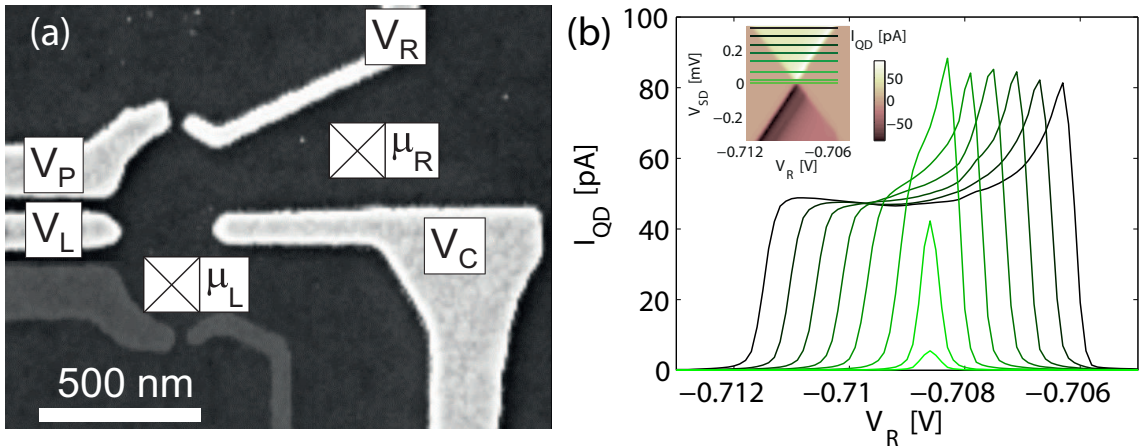}
 \caption{\label{figure_1}(a) Scanning electron micrograph of the relevant part of the sample. The dark grey area is the surface of the GaAs/AlGaAs heterostructure, the bright regions correspond to the Ti/Au top gates which were used for the experiment. The gates which were kept on ground during the whole experiment are greyed out. A DC-bias voltage was applied between the left ($\mu_\mathrm{L}$) and right ($\mu_\mathrm{R}$) contact. (b) The measured current through the quantum dot for different biases in a perpendicular magnetic field $B=3$~T (see inset for corresponding line cuts in the Coulomb Diamond). The Fermi edge singularity develops once the bias $\left|\mu_\mathrm{L}-\mu_\mathrm{R}\right|$ exceeds the level width, and becomes more pronounced with further increase of the bias.}
\end{figure}

The measured samples were fabricated using a high-mobility GaAs/AlGaAs heterostructure with a two-dimensional electron gas (2DEG) $90$~nm below the surface. The electronic mobility was $\mu_\mathrm{e}=2.2 \times 10^6$~cm$^2$/Vs at $T=1.3$~K and the electron density $n_\mathrm{s}=2.0\times10^{11}$~cm$^{-2}$. Quantum dots (QDs) were formed and tuned into the Coulomb blockade regime by applying negative voltages to Ti/Au Schottky gates deposited by standard electron beam lithography. We observed the effects reported here in several samples with different gate layouts. Here we only present data of the sample shown in figure~\ref{figure_1}(a), which has been studied in more detail. However, the conclusions are also valid for all the other devices. Only the four bright gates shown in figure~\ref{figure_1}(a) were used for this experiment. The two lower gates (coloured darker) which enable the formation of a double quantum dot were kept on ground during all measurements. We apply a DC-bias $-eV_\mathrm{SD}=\mu_\mathrm{L}-\mu_\mathrm{R}$ between the left and right contact ($\mu_\mathrm{L,R}$ denote the electrochemical potential of the left and right lead respectively) and measure the resulting DC-current with a standard current to voltage converter.
All measurements were performed in a $^3$He/$^4$He dilution refrigerator with the electronic base temperature $T_\mathrm{el}=25$~mK extracted from tunnelling resonance peak widths in the Coulomb blockade regime \cite{beenakker_theory_1991}. A magnetic field of $B=4$~T ($B=3$~T for figure \ref{figure_1}(b)) is applied perpendicular to the plane of the 2DEG forming quantised Landau levels. We measure the filling factor $\nu=2$ ($\nu=3$) plateau for a magnetic field $B=4$~T ($B=3$~T). The Fermi edge singularity is also visible at zero magnetic field, but gets more prominent for higher magnetic fields, similar to previous reports \cite{hapke-wurst_magnetic-field-induced_2000,frahm_fermi-edge_2006}.\\
Due to an impurity near the quantum dot, sweeping the plunger gate ($V_\mathrm{P}$) resulted in noisy traces and this voltage had to be kept constant during the measurements. As a consequence the right tunnelling barrier gate ($V_\mathrm{R}$) serves two purposes: while being swept over a small range it acts mostly as a plunger and influences the electrochemical potential $\mu_\mathrm{QD}$ of the quantum dot. If changed by a large amount it changes the relative strength of the tunnelling barrier.
\\
\section{Results}
In figure \ref{figure_1}(b) the current through the quantum dot at $B=3$~T is plotted for different bias voltages $V_\mathrm{SD}$ (see inset) applied between the left and right reservoir as a function of gate voltage $V_\mathrm{R}$. For low bias voltages the current shows a symmetric peak, which follows strictly neither the temperature broadened nor the lifetime broadened resonance. This indicates, as will be also shown later, that the lifetime broadening and the smearing due to temperature are of the same order of magnitude. For higher biases the line shape clearly becomes asymmetric and has a pronounced resonance at the high-voltage edge (see inset for the Coulomb diamond measurement from which the line traces were extracted). This resonant enhancement occurs when $\mu_\mathrm{QD}$ is in resonance with the Fermi energy of the left lead and can be identified as a Fermi edge singularity. The magnitude of the resonance is independent of the applied bias voltage, i.e. the current depends on $\mu_\mathrm{L}-\mu_\mathrm{QD}$ but not on $\mu_\mathrm{R}$. In contrast to the vertical tunnelling devices the bias voltage applied here is only needed to separate the right and left Fermi reservoirs sufficiently far from each other in energy, such that the resonance becomes discernible.
\begin{figure}[t!hb!]
 \centering
 \includegraphics[width=\textwidth]{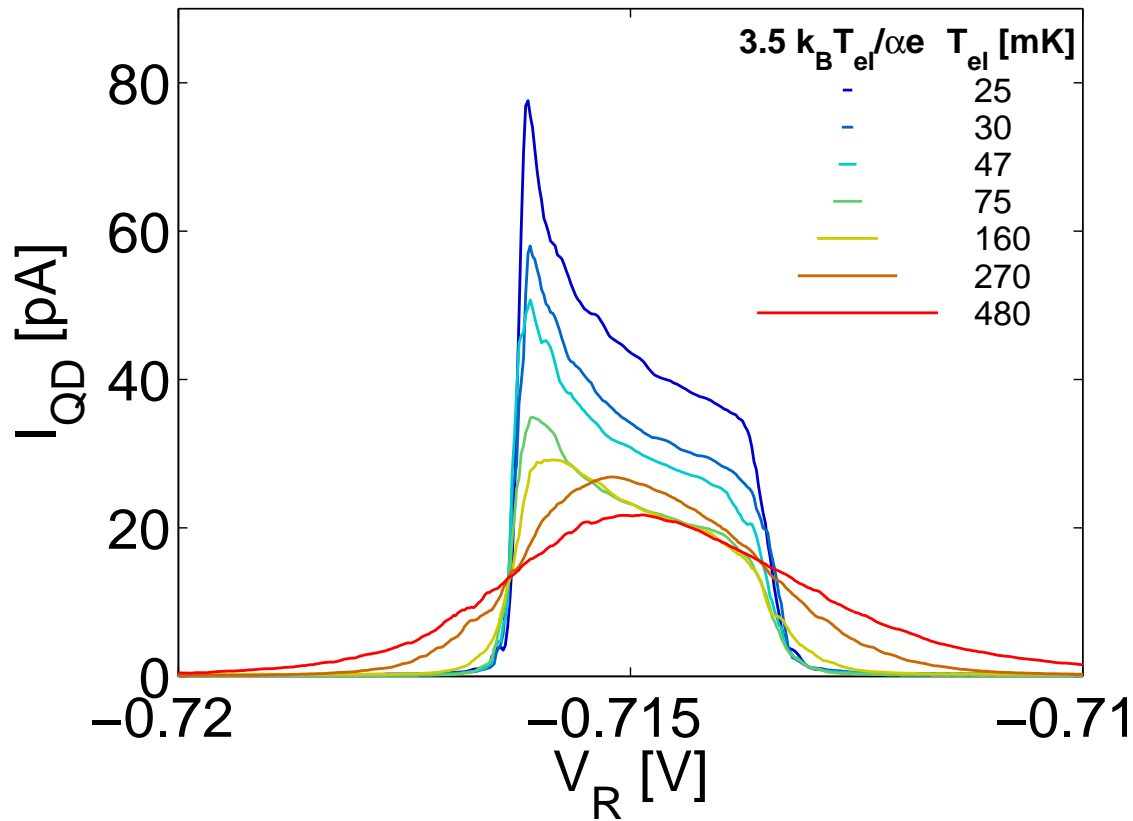}
 \caption{\label{figure_2}Current through the quantum dot for different electronic temperatures in a perpendicular field $B=4$~T. $V_\mathrm{R}$ serves as a plunger gate voltage and influences $\mu_\mathrm{QD}$. A constant DC-bias of 200~$\mu$V was applied. The temperature was varied from $T_\mathrm{el}=25$~mK (blue) to $T_\mathrm{el}=480$~mK (red) (extracted from Coulomb peak widths at zero bias). The thermal broadening of the Fermi reservoirs (roughly $3.5\times k_\mathrm{B}T_\mathrm{el}$) converted to gate voltage is indicated by the horizontal lines on the top right part.}
\end{figure}\\
The Fermi edge singularity is very susceptible to temperature changes. The resonance shape is expected to change as soon as $k_\mathrm{B}T_\mathrm{el}$ exceeds the broadening due to the finite lifetime. Figure~\ref{figure_2} shows the measured temperature dependence of the Fermi edge singularity. The horizontal lines on the top right indicate the thermal broadening in gate voltage (3.5$\times k_\mathrm{B}T_\mathrm{el}/e\alpha$, where $\alpha$ is the leverarm of gate $V_\mathrm{R}$ on the quantum dot, relating a change of voltage on $V_\mathrm{R}$ to an energy change of the quantum dot: $\alpha=\Delta\mu_\mathrm{QD}/e\Delta V_\mathrm{R}$). The temperature is extracted from a Coulomb resonance peak width in the weak coupling limit and converted with the leverarm determined from a Coulomb diamond measurement. As the temperature smearing and the lifetime broadening are of the same order of magnitude at base temperature we already see a decrease of the resonance for the slightest change of temperature. At moderate electronic temperatures of $75$~mK the resonance is barely discernible, indicating the small energy scale and fragile interaction causing the effect. This temperature  is much smaller than those reported previously in vertical tunnelling structures, where the resonance was observed up to several Kelvin \cite{hapke-wurst_magnetic-field-induced_2000,geim_fermi-edge_1994}. This might be attributed to the larger distance between the Fermi sea and the localised state in our system, compared to a vertical alignment. For the highest temperatures of 480~mK reported here thermal broadening becomes as large as the applied bias voltage and thus decreases the overall amplitude of the current.\\
The main advantage of our system compared to vertical devices is the tunability of tunnelling barriers. Figure~\ref{figure_3} shows the Fermi edge singularity for different tunnelling barrier strengths at base temperature. We change from a situation where the right tunnelling coupling is much weaker than the left (see figure~\ref{figure_3}(a)) to the opposite situation (figure~\ref{figure_3}(f)). Likewise the Fermi edge singularity shifts from the "right" side of the bias window to the "left" side. Thus the resonance always occurs when the quantum dot electrochemical potential is aligned with the Fermi energy of the reservoir which has the weaker tunnelling coupling, in agreement with theory \cite{matveev_interaction-induced_1992}. The reason is that the current is dominated by the higher tunnelling barrier. This dependence is universally observed in all samples. We want to point out that the applied bias and in particular its direction has not been changed in figures.~\ref{figure_3}~(a-f). This means that we change from a situation where the electron is mostly on the dot and tunnels from it (figures~\ref{figure_3} (a-c)) to a situation where the dot is mostly empty and is filled by the tunnelling electron (figures~\ref{figure_3} (d-f)), indicating the particle hole-symmetry of this process.
\begin{figure}[t!hb!]
 \centering
 \includegraphics[width=\textwidth]{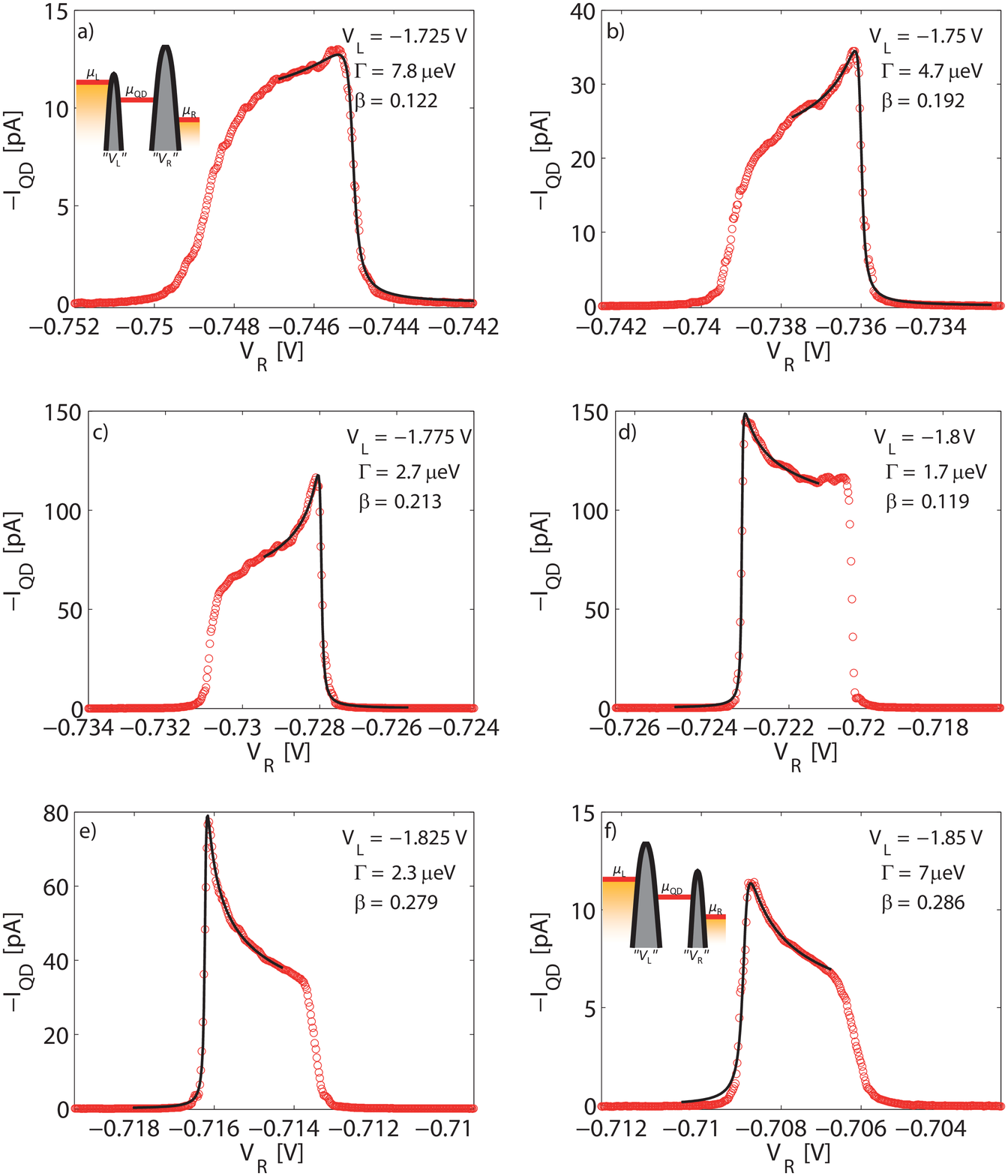}
 \caption{\label{figure_3}(a)-(f) Current through the quantum dot for different tunnelling barrier strengths (red circles) at base temperature. A constant dc-bias ($-200~\mu$V) and a perpendicular magnetic field ($B=4$~T) were applied. The plunger gate voltage $V_\mathrm{P}$ and the center gate voltage $V_\mathrm{C}$ were kept constant while the two barrier gates $V_\mathrm{L,R}$ were changed. While for the small scan range of each panel the voltage $V_\mathrm{R}$ acts as a plunger gate (shifting $\mu_\mathrm{QD}$) changing both barrier voltages from (a) to (f) changes the relative strength of the tunnelling barriers drastically (see inset for schematics). Each trace was fitted with formula~\ref{formula_1} (black trace) and the relevant fit parameters $\Gamma$ and $\beta$ are displayed together with the voltage applied to $V_\mathrm{L}$. Note: panel (e) corresponds to the lowest temperature trace of figure~\ref{figure_2}.}
\end{figure}\\
Figure~\ref{figure_3} shows in black the fit to the data, following (\ref{formula_1}). Strictly speaking (\ref{formula_1}) assumes the zero-temperature limit which is not the case for our situation. Nevertheless fitting the curves yields excellent agreement between theory and data, except in the tails (see for instance figure~\ref{figure_3} (f)) which is most likely the effect of the non-negligible temperature broadening. The relevant fit parameters $\Gamma$ and $\beta$ are shown for the individual fits in the figures. We want to emphasise here again, that the singularity is due to the interaction with the Fermi reservoir which is less coupled but the smearing of the singularity will be due to finite lifetime of the state which is governed by the more strongly coupled reservoir. Experimentally we can not distinguish between the broadening due to tunnelling from the left and right side, thus we can only extract the overall broadening $\Gamma$. The values of $\Gamma$ of the order of a few~$\mu$eV are consistent with a life-time broadened Coulomb peak for these current values. Even the dependence on the tunnelling barrier strength follows the expected tendency as we would expect the longest lifetime (smallest $\Gamma$) for a symmetric tunnelling coupling (i.e. figures~\ref{figure_3}~(c)~and~(d)).
However, changing the value of $V_\mathrm{P}$ slightly and repeating the measurement yields quite different parameter values. We do not know the origin of these differences. This also prevents us from extracting a quantitative dependence on magnetic field (as done in \cite{hapke-wurst_magnetic-field-induced_2000}) because the magnetic field changes the transmission of the tunnelling barriers. Thus, for each value of magnetic field the values of $V_\mathrm{R}$,$V_\mathrm{L}$ and $V_\mathrm{P}$ had to be adjusted. In general we find that a higher magnetic field makes the resonance more discernible. Summarizing the analysis we can say that the qualitative features of our experimental observations are well explained by theory.

\section{Conclusion}

We reported the observation of a Fermi edge singularity in lateral tunnelling through a quantum dot. We have shown the characteristic temperature dependence associated with the Fermi edge singularity. The resonance is dominated by the higher tunnelling barrier, limiting the overall current. The position of the resonance can be changed from the upper edge to the lower edge of the bias window by swapping the more resistive tunnelling barrier from the negative to the positive pole.

\section{Acknowledgements}

We would like to acknowledge fruitful discussions  with Yuval Gefen, Leonid Levitov and Yashar Komijani. We gratefully acknowledge the support of the ETH FIRST laboratory, the financial support of the Swiss National Science Foundation (Schweizerischer Nationalfonds, NCCR QSIT) and NSF Grant DMR-1603243.
\clearpage

\section*{References}
\bibliography{literature}
\bibliographystyle{iopart-num}
%\begin{thebibliography}{litshort}
%\bibitem{ref1} J.~Doe, Article name, \textit{Phys. Rev. Lett.}
%\bibitem{ref2} J.~Doe, J. Smith, Other article name, \textit{Phys. Rev. Lett.}
%\bibitem{web} \href{http://www.google.pl}{www.google.pl}
%\end{thebibliography}

\end{document}